\documentclass[12pt]{article}

\usepackage[english]{babel}
\usepackage[utf8]{inputenc}
\usepackage{amsmath,amsthm,amsfonts,amscd,amssymb,eucal,latexsym,slashed}
\usepackage{cite}
\usepackage{epsfig}
\usepackage{graphicx}

\newtheorem{theorem}{Theorem}[section]
\newtheorem{definition}[theorem]{Definition}

\newtheorem{lemma}[theorem]{Lemma}
\newtheorem{proposition}[theorem]{Proposition}

\newcommand{\ie}{{\it i.e.\ }}
\newcommand{\eg}{{\it e.g.\ }}

\def\RR{{\mathbb R}}
\def\RRO{{\mathbb R} \backslash \{0\}}

\def\NN{{\mathbb N}}

\def\ZZ{{\mathbb Z}}

\def\Hs{{\cal H}}
\def\Is{{\cal I}}

\def\Ks{{\cal K}}

\def\Ps{{\cal P}}

\def\Rs{{\cal R}}

\def\Ws{{\cal W}}

\newcommand{\Xsig}{(X, \sigma)}

\newcommand{\PA}{{\Ps (X, \sigma)}}  %  polynomial algebra
\newcommand{\WA}{{\Ws (X, \sigma)}}  %  Weyl algebra
\newcommand{\RAO}{{\Rs_0 (X, \sigma)}}  %  pre--Resolvent algebra
\newcommand{\RA}{{\Rs (X, \sigma)}}  %  Resolvent algebra

\newcommand{\Rlf}{{R(\lambda,f)}}
\newcommand{\Rmg}{{R(\mu,g)}}

\newcommand{\un}{{\bf 1}}

\newcommand{\Ad}[1]{(\mbox{\rm Ad} \, #1 \, )}

\newcommand{\Rep}{{(\pi, \Hs)}}

\newcommand{\rest}{\upharpoonright}

\title{Quantum Systems and Resolvent Algebras}
\author{{}\\[-4mm] {\Large Detlev Buchholz\,$^a$\thanks{Supported by
the German Research Foundation (Deutsche Forschungsgemeinschaft
(DFG)) through the Institutional Strategy of the University of G\"ottingen}
 \ and \    Hendrik Grundling\,$^b$}\\[10mm]
${}^a$ Institut f\"ur Theoretische Physik and
Courant Centre \\ ``Higher Order Structures in Mathematics'',
Universit\"at G\"ottingen, \\ 37077 G\"ottingen, Germany  \\[2mm]
${}^b$   Department of Mathematics,  University of New South Wales \\
Sydney, NSW 2052, Australia}

\date{}

\begin{document}

\maketitle

\begin{abstract}
\noindent
This survey article is concerned with the modeling of the
kinematical structure of quantum systems 
in an algebraic framework which eliminates certain 
conceptual and computational difficulties of the conventional
approaches. Relying on the Heisenberg picture
it is based on the resolvents of the basic canonically
conjugate operators and
covers finite and infinite quantum systems. The resulting
C*--algebras, the resolvent algebras,
have many desirable properties. On one hand they encode
specific information about the dimension of the respective
quantum system and have the mathematically comfortable
feature of being nuclear, and for finite dimensional systems
they are even postliminal. This comes along with a surprisingly simple
structure of their representations. On the other hand,
they are a convenient framework for the study of interacting
as well as constrained quantum systems since they allow
the direct application of C*--algebraic methods which often
simplify the analysis. Some pertinent facts are illustrated
by instructive examples.
\end{abstract}

\newpage

\section{Introduction}
\setcounter{equation}{0}

The conceptual backbone for the modeling of
the kinematics of quantum
systems is the Heisenberg commutation relations which have found
their mathematical expression in various guises. There is an
extensive literature analyzing their properties, starting
with the seminal paper of Born, Jordan and Heisenberg on the
physical foundations and reaching a first mathematical satisfactory
formulation in the works of von Neumann and of Weyl.

These canonical systems of operators may all be presented in
the following general form: there is a real
(finite or infinite dimensional) vector space $X$
equipped with a non--degenerate symplectic form
$\sigma : X \times X \rightarrow \RR$ and a linear map $\phi$  from $X$
onto the generators of a polynomial *--algebra $\PA$ of
operators satisfying the canonical commutation relations
$$
\big[\phi(f),\,\phi(g)\big]=i\sigma(f,\, g) \, {\un}, \quad
\phi(f)^* =\phi(f) \, .
$$
In the case that $X$ is finite dimensional, one can reinterpret
this relation in terms of the familiar quantum mechanical
position and momentum operators, and if $X$ consists
of Schwartz functions on some
manifold one may consider $\phi$ to be a bosonic quantum field.
As is well--known, the operators $\phi(f)$ cannot all be bounded.
Moreover, the algebra  $\PA$  does not admit 
much interesting dynamics acting on it by automorphisms; in
fact there are in general only transformations  
induced by polynomial Hamiltonians 
which leave it invariant \cite{Di}. Thus  
$\PA$ is not a convenient kinematical algebra in either
respect.

The inconveniences of unbounded operators 
can be evaded by expressing the basic commutation relations
in terms of bounded functions of the generators $\phi(f)$.
In the approach introduced by Weyl, this is done by
considering the C*--algebra generated by the set of
unitaries  $W(f) \, \hat{=} \, \exp(i\phi(f))$, $f \in X$
(the Weyl operators) satisfying the Weyl relations
$$
W(f) W(g) =
e^{-i \sigma(f,g)/2} \, W(f+g) \, , \quad
W(f)^* = W(-f) \, .
$$
This is the familiar Weyl (or CCR) algebra $\WA$. Yet
this  algebra still suffers from the fact that its automorphism
group does not contain physically significant dynamics \cite{FaVB}.
This deficiency can be traced back to the
fact that the Weyl algebra is simple, whereas any unital
C*--algebra admitting an expedient variety of dynamics must
have ideals \mbox{\cite[Sec.~10]{BuGr1}}, cf.\ also the conclusions.

For finite systems this problem can be solved by 
proceeding to the twisted group algebra~\cite{Ka} 
derived from the unitaries $W(f)$, $f \in X$. 
By the Stone--von Neumann theorem this
algebra is isomorphic to $\Ks(\Hs)$, the compact operators 
on a separable Hilbert space, for any finite dimensional $X$. 
This step solves the problem of dynamics for finite systems, but 
it cannot be applied as such to infinite systems
since there $X$ is not locally compact. Moreover, one pays
the price that the original operators, having continuous spectrum,
are not affiliated with $\Ks(\Hs)$. So 
one forgets the specific properties of the underlying quantum 
system.

This unsatisfactory situation motivated the formulation of
an alternative version of the C*--algebra of canonical commutation
relations, given in \cite{BuGr1}. Here one considers
the C*--algebra generated by the resolvents of the
basic canonical operators which are formally given by
$\Rlf \, \hat{=} \, (i \lambda \un - \phi(f))^{-1}$
for $\lambda \in \RRO$, $f \in X$. All algebraic
properties of the operators~$\phi(f)$ can be expressed in terms of
polynomial relations amongst these resolvents. Hence,
in analogy to the Weyl algebra generated by the exponentials,
one can abstractly define a unital \mbox{C*--algebra}
$\RA$ generated by the resolvents, called the resolvent
algebra.

In accordance with the requirement of admitting sufficient dynamics
the resolvent algebras have ideals. Their ideal
structure was recently clarified in \cite{BuGr2}, where it was shown
that it depends sensitively on the size of the underlying
quantum system. More precisely, the specific nesting of the
primitive ideals encodes information about the dimension of
the underlying space $X$. This dimension,
if it is finite, is an algebraic invariant which labels the
isomorphism classes of the resolvent algebras.
Moreover, the primitive ideals are in one--to--one
correspondence to the spectrum (dual) of the respective
algebra, akin to the case of commutative algebras.
The resolvent algebras are postliminal (type I) if the
dimension of $X$ is finite and they are still nuclear if
$X$ is infinite dimensional. Thus these algebras not only encode
specific information about the underlying systems but
also have comfortable mathematical properties.

The resolvent algebras already have proved to be useful in several
applications to quantum physics such as the representation theory of abelian
Lie algebras of derivations \cite{BuGr3}, the study
of constraint systems and of the BRST method in a
C*--algebraic setting \cite{BuGr1,Co}, the
treatment of supersymmetric
models on non--compact spacetimes and the rigorous construction
of corresponding JLOK--cocycles \cite{BuGr4}. Their 
virtues also came to light in the formulation and 
analysis of the dynamics of finite
and infinite quantum systems \cite{BuGr1,BuGr5}.

In the present article we give a survey of the basic
properties of the resolvent algebras and an outline of
recent progress in the construction of dynamics,
shedding light on the role of the ideals.
The subsequent section contains the formal definition
of the resolvent algebras and some comments on their relation
to the standard Weyl formulation of the canonical
commutation relations. Section~3 provides a
synopsis of representations of the resolvent algebras
and some structural implications and Sect.~4 contains the
discussion of observables and of dynamics. The article
concludes with a brief summary and outlook.

\section{Definitions and basic facts}
\setcounter{equation}{0}

Let $\Xsig$ be a real symplectic space;
in order to avoid pathologies we make the standing assumption
that $\Xsig$ admits a unitary structure \cite{Ro}. 
The pre--resolvent algebra $\RAO$ is the universal \mbox{*--algebra}
generated by the elements of the set
$\{\Rlf : \lambda \in \RRO, \; f\in X \}$
satisfying the relations
\begin{eqnarray}
\label{Resolv}
\Rlf - R(\mu,f) & \! \! = \! \! & i (\mu-\lambda) \Rlf R(\mu,f)  \\[1mm]
\label{Rinvol}
\Rlf^*& \! \! = \! \! &R(-\lambda,f) \\[1mm]
\label{Rccr}
\big[\Rlf, \, \Rmg \big] & \! \! = \! \! &
i\sigma(f,g)\,\Rlf \, \Rmg^2 \Rlf \\[1mm]
\label{Rhomog}
\nu \,R(\nu \lambda,\, \nu f)   & \! \! = \! \! &  \Rlf \\[1mm]
\label{Rsum}
\Rlf \Rmg & \! \! = \! \! & R(\lambda+\mu, f+g) \big( \Rlf + \Rmg
+ i\sigma(f,g) \Rlf^2 \Rmg \big) \\[1mm]
\label{Riden}
R(\lambda,0)& \! \! = \! \! & -{\textstyle \frac{i}{\lambda}} \, \un
\end{eqnarray}
where $\lambda,\, \mu, \, \nu \in \RRO $ and $f,\,g\in X$,
and for (\ref{Rsum}) we require $\lambda+\mu\not=0$.
That is, start with the free unital *-algebra generated by
$\{\Rlf : \lambda \in \RRO, \; f\in X \}$
 and factor out by the ideal generated by the relations
(\ref{Resolv}) to (\ref{Riden}) to obtain the *-algebra $\RAO$.

\medskip
\noindent \textbf{Remarks:}
(a) Relations (\ref{Resolv}), (\ref{Rinvol}) encode the
algebraic properties of the resolvent of some self--adjoint operator,
(\ref{Rccr}) amounts to the canonical commutation relations and relations
(\ref{Rhomog}) to (\ref{Riden}) correspond to the linearity of the 
initial map $\phi$ on $X$. \\
(b) \ The *-algebra $\RAO$ is nontrivial, because it has nontrivial
representations. {}For instance, in a Fock representation
$\Rep$ one has self--adjoint operators $\phi_\pi (f)$, $f \in X$
satisfying the canonical commutation relations over $\Xsig$
on a sufficiently big domain in the Hilbert space $\Hs$
so that one can define $\pi(\Rlf) \doteq (i\lambda\un - \phi_\pi (f))^{-1}$
to obtain a representation~$\pi$ of~$\RAO$.

\medskip
It has been shown in  \cite[Prop.~3.3]{BuGr1} that the following
definition is meaningful.

\begin{definition} Let $\Xsig$  be a symplectic space.
The supremum of operator norms with regard to all
cyclic *--representations $\Rep$  of  $\, \RAO$
$$ \| R \| \doteq \sup_\Rep \, \| \pi(R) \|_\Hs \, , \quad R \in \RAO  $$
exists and defines a C*--seminorm on $\RAO$.
The resolvent algebra $\RA$ is defined as the C*--completion
of the quotient algebra $\RAO / \ker \| \cdot \|$,
where here and in the following the symbol $\ker$ denotes
the kernel of the respective map.
\end{definition}

Of particular interest are representations of the resolvent
algebras, such as the Fock representations, where the abstract
resolvents characterized by conditions (\ref{Resolv}), (\ref{Rinvol})
(sometimes called pseudo--resolvents)
are represented by genuine resolvents of self--adjoint operators.

\begin{definition}
A representation $\Rep$ of $\RA$ is said to be regular if
for each $f \in X$ there exists a self--adjoint operator 
$\phi_\pi(f)$ such that $\pi(\Rlf) \doteq (i\lambda\un - \phi_\pi
(f))^{-1}$, $\lambda \in \RRO$. (This is equivalent to the condition
that all operators $\pi(\Rlf)$ have trivial kernel.)
\end{definition}

The following result characterizing regular representations, cf.\
\cite[Thm.~4.10 and Prop.~4.5]{BuGr1}, is of importance, both in the
structural analysis of the resolvent algebras and in their
applications. It implies in particular that the resolvent algebras
have faithful irreducible representations (\eg the
Fock representations), so their centers are trivial.

\begin{proposition} \label{faithfulrep}
Let $\Rep$ be a representation of $\RA$.
\begin{itemize}
\vspace*{-2mm}
\item[(a)] If $\Rep$ is regular it is also faithful,
\ie $\| \pi(R) \|_\Hs = \| R \| $ for $R \in \RA$.
\vspace*{-2mm}
\item[(b)] If $\Rep$ is faithful and
the weak closure of $\pi(\RA)$ is a factor, then $\Rep$ is regular.
\end{itemize}
\end{proposition}

The regular representations of the resolvent algebras are in
one--to--one correspondence with the regular representations
of the Weyl--algebras, cf. \cite[Cor.~4.4]{BuGr1}. 
(Recall that a representation $\Rep$ of $\WA$ 
is regular if the maps $\nu \in \RR\mapsto\pi(W(\nu f))$ 
are strong operator 
continuous for all $f$.) In fact one has the following result.

\begin{proposition} \label{one-one}
Let $\Xsig$ be a symplectic space and
\begin{itemize}
\vspace*{-2mm}
\item[(a)] let $\Rep$ be a regular representation of
the resolvent algebra $\RA$ with associated self--adjoint operators
$\phi_\pi(f)$ defined above. The exponentials
$W_\pi(f) \doteq \exp(i \phi_\pi(f))$, $f \in X$ satisfy the Weyl relations
and thus define a regular representation of the Weyl algebra~$\WA$
on~$\Hs$;
\vspace*{-2mm}
\item[(b)] let $\Rep$ be a regular representation of
the Weyl algebra $\WA$ and let $\phi_\pi(f)$ be the 
self--adjoint generators of the Weyl operators. The resolvents
\mbox{$R_\pi(\lambda,f) = (i\lambda\un - \phi_\pi (f))^{-1}$} 
with $\lambda \in \RRO$,
$f \in X$ satisfy relations (\ref{Resolv}) to (\ref{Riden}) and
thus define a regular representation of the resolvent algebra
$\RA$ on $\Hs$.
\end{itemize}
\end{proposition}

Whilst this proposition establishes the
existence of a bijection between the regular
representations of $\RA$ and those of $\WA$, there is no such
map between the non--regular representations of the two
algebras. In order to substantiate this point
consider for fixed nonzero $f \in X$  the  two commutative subalgebras
$C^*\{ R(1,sf) : s \in \RR \} \subset \RA$
and $C^*\{ W(sf) : s \in \RR \} \subset \WA$.
These algebras are isomorphic
respectively to the continuous functions on
the one point compactification of $\RR$, and the continuous functions on
the Bohr compactification of $\RR$. Now the point measures
on the compactifications having support in the complement of $\RR$
produce non-regular states (after extending to the full C*--algebras by
 Hahn--Banach) and there are many more of these for
the Bohr compactification than for the one point compactification of
$\RR$. Proceeding to the GNS--representations it is
apparent that the Weyl algebra has substantially more
non-regular representations than the resolvent~algebra.

\section{Ideals and dimension}
\setcounter{equation}{0}

Further insight into the algebraic properties of the resolvent
algebras is obtained by a study of its irreducible representations.
In case of finite dimensional symplectic spaces these representations
have been completely classified \cite[Prop.~4.7]{BuGr1}.

\begin{theorem}
Let $\Xsig$ be a finite dimensional symplectic space and let
$\Rep$ be an irreducible representation of $\RA$. Depending
on the representation, the space $X$ decomposes as follows, 
cf.\ Fig.\ 1.
\begin{itemize}
\vspace*{-2mm}
\item[(a)] There is a unique subspace $X_R \subset X$ such that
there are self--adjoint operators $\phi_\pi(f_R)$ satisfying
$\pi(R(\lambda,f_R)) =   (i\lambda\un - \phi_\pi (f_R))^{-1}$ for
$\lambda \in \RRO$, $f_R \in X_R$.
\vspace*{-2mm}
\item[(b)]
Let $X_T \doteq 
\{f\in X_R : \sigma(f,g)=0 \ \mbox{for all} \ g\in X_R \}$.
Then $\phi_\pi$  restricts on $X_T$ to a linear functional 
$\varphi : X_T \rightarrow \RR$ such that
$\pi(R(\lambda,f_T)) =   (i\lambda - \varphi (f_T))^{-1} \un$
for $f_T \in X_T$, $\lambda \in \RRO$.
\vspace*{-2mm}
\item[(c)] For $f_S \in X_S \doteq X \backslash X_R$ and $\lambda \in
  \RRO$ one has $\pi(R(\lambda,f_S)) = 0$.
\end{itemize}
\vspace*{-2mm}
Conversely, given subspaces $X_T \subset X_R \subset X$ and
a linear functional $\varphi : X_T \rightarrow \RR$ there exists a
corresponding irreducible representation $\Rep$ of $\RA$,
unique up to equivalence, with the preceding three properties.
\end{theorem}

\begin{figure}[h] \label{figure}
 \hspace*{60mm}
\epsfig{file=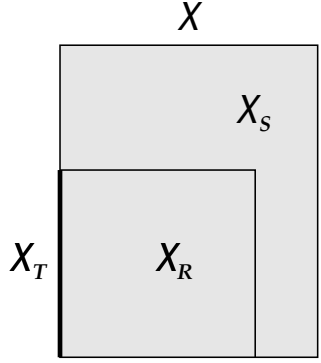,height=48mm}
\caption{Decomposition of $X$ fixed by an
irreducible representation}
 \label{fig1}
 \end{figure}

This result may be regarded as an extension of the
Stone--von Neumann uniqueness theorem for regular representations
of the CCR algebra. It shows that the only obstruction to
regularity is the possibility that some of the underlying
canonical operators are infinite and the corresponding
resolvents vanish. This happens in
particular if there are some canonically conjugate operators
having sharp (non--fluctuating) values in a representation, as is the case for
constraint systems \cite[Prop.~8.1]{BuGr1}.
But, in contrast to the Weyl algebra,
the non--regular representations of the resolvent algebra
only depend on the values of these canonical operators. So
the abundance of different singular representations of the Weyl 
algebra shrinks to a manageable family on the resolvent algebra.

The preceding theorem is the key to the structural analysis
of the resolvent algebra for symplectic spaces of arbitrary
finite dimension. We recall in this context that the
primitive ideals of a C*--algebra are the (possibly zero)
kernels of irreducible representations and that the spectrum of the algebra
is the
set of unitary equivalence classes of irreducible representations.
The following
result has been established in \cite{BuGr2}.

\begin{theorem}
Let $\Xsig$ be a finite dimensional symplectic space.
\begin{itemize}
\vspace*{-2mm}
\item[(a)] The mapping $\widehat{\pi} \mapsto \ker \widehat{\pi}$
from the elements $\widehat{\pi}$ of the
spectrum (dual) of the resolvent algebra $\RA$ to its primitive
ideals \ $\ker \widehat{\pi}$ is a bijection.
\vspace*{-2mm}
\item[(b)] Let $L \doteq
\sup \, \{l \in \NN : \ker \widehat{\pi}_1 \subset
\ker \widehat{\pi}_2 \dots \subset \ker \widehat{\pi}_l \}$
be the maximal length of proper inclusions of primitive ideals
of $\RA$. Then $L  = \dim(X) / 2 + 1$.
\end{itemize}
\end{theorem}

\noindent \textbf{Remarks:}
Property (a) is a remarkable feature of the resolvent algebras,
shared with the abelian \mbox{C*--algebras}. It rarely holds for
non-commutative algebras and also fails if $X$ is infinite dimensional.
The quantity $L$ defined in (b) is an algebraic invariant,
so this result shows that the dimension $\dim(X)$ of the underlying systems
is algebraically encoded in the resolvent algebras. As a matter of fact,
$\dim(X)$ is a complete algebraic invariant of resolvent algebras
in the finite dimensional case.

\vspace*{2mm}
As indicated above, there is an algebraic difference between the resolvent
algebras for finite dimensional $X$ and those where $X$ has infinite dimension.
A further difference is seen through the minimal (nonzero) ideals \cite{BuGr2}.

\begin{proposition} \label{compactideal}
Let $\Xsig$ be a symplectic space of arbitrary dimension and
let $\Is \subset \RA$ be the intersection of all nonzero ideals of $\RA$.
\begin{itemize}
\vspace*{-2mm}
\item[(a)] If \ $\dim(X) < \infty$ then $\Is$ is isomorphic to the
C*--algebra $\Ks(\Hs)$ of compact operators. 
Moreover, in any irreducible regular representation $\Rep$ one has 
$\pi(\Is)=\Ks(\Hs)$.
\vspace*{-2mm}
\item[(b)] If  \ $\dim(X) = \infty$ then $\Is = \{0\}$. In fact,
there exists no nonzero minimal ideal of $\RA$ in this case.
\end{itemize}
\end{proposition}

If $\Xsig$ is infinite dimensional the
resolvent algebra $\RA$ is the \mbox{C*--inductive} limit of
the net of its subalgebras $\Rs(Y,\sigma)$ where
$Y \subset X$ ranges over all finite dimensional
non--degenerate subspaces of $X$, cf.~\cite[Thm.~4.9]{BuGr1}.
This fact in combination with the first part of the preceding result
is a key ingredient in the construction of
dynamics, see below. It also enters in the proof
of the following statement \cite{BuGr2}.

\begin{proposition}
Let $(X,\sigma)$ be a symplectic space of arbitrary
dimension.
\begin{itemize}
\vspace*{-2mm}
\item[(a)] $\RA$ is a nuclear C*-algebra,
\vspace*{-2mm}
\item[(ii)] $\RA$ is a postliminal (type~I) C*--algebra
if and only if  $\dim(X)<\infty$.
\end{itemize}
\end{proposition}

Recall that a C*--algebra is said to be postliminal (type~I) if
all of its irreducible representations contain the
compact operators and that postliminal C*--algebras
as well as their C*--inductive limits are nuclear,
\ie their tensor product with any other C*--algebra
is unique. It should be noted, however, that the
resolvent algebras are not separable \cite[Thm.~5.3]{BuGr1}.
With this remark we conclude our outline of pertinent
algebraic properties of the resolvent algebras.

\section{Observables and dynamics}
\setcounter{equation}{0}

The main virtue of the resolvent algebra consists of the fact
that it includes many observables of physical  interest and
admits non--trivial dynamics. In order to illustrate this
important feature we discuss in detail a familiar example of a  
finite quantum system and comment on infinite systems at the 
end of this section.

Let $(X,\sigma)$ be a finite dimensional symplectic space, \ie
$\dim(X) = 2N$ for some \mbox{$N \in \NN$}. Since
regular representations of the resolvent algebras
are faithful, cf.\ Proposition~\ref{faithfulrep},
it suffices to consider a regular irreducible representation
$(\pi_0,\Hs_0)$ of $\Rs(X,\sigma)$ which is unique up to
equivalence.
Choosing some symplectic basis $f_k, g_k \in X$ and putting
\mbox{$P_k \doteq \phi_{\pi_0}(f_k)$},  $Q_k \doteq  \phi_{\pi_0}(g_k)$,
\mbox{$k = 1, \dots N$} we identify the self--adjoint operators
fixed by the corresponding resolvents with the momentum and
position operators of~$N$ particles in one spatial dimension.

The (self--adjoint) quadratic Hamiltonian
$$
H_{0} \doteq \sum_{k=1}^N \, ({\textstyle \frac{1}{2m_k}} P_k^2 +
{\textstyle \frac{m_k \omega_k^2}{2}} Q_k^2)
$$
describes the free, respectively oscillatory motion of these particles,
where $m_k$ are the particle masses and $\omega_k \geq 0$ the
frequencies of oscillation, $k = 1, \dots N$. The interaction of the
particles is described by the operator
$$
V \doteq \sum_{1 \leq k < l \leq N} \, V_{k l}(Q_k - Q_l)
$$
where we assume for simplicity that the potentials  $V_{k l}$
are real and continuous, vanish at infinity, but are
arbitrary otherwise. Since $V$ is bounded, the
Hamiltonian $H \doteq H_{0} + V$
is self--adjoint on the domain of $H_{0}$ and its resolvents
are well defined.

\begin{proposition} \label{observables}
Let $H$ be the Hamiltonian defined above. Then
$$ (i \mu \un - H)^{-1} \in \pi_0(\Rs(X,\sigma)) \, ,
\quad \mu \in \RRO \, . $$
\end{proposition}

\vspace*{2mm}
\noindent \textbf{Remark:} Since $\pi_0$ is faithful
its inverse $\pi_0^{-1} : \pi_0(\RA) \rightarrow
\RA$ exists,
so this result shows that $H$ is affiliated with the
resolvent algebra. Note that this is neither true
for the Weyl algebra $\WA$ nor for the corresponding twisted 
group algebra $\Ks(\Hs)$ if one of the frequencies~$\omega_k$
vanishes. Thus $\RA$ contains many more observables
of physical interest than these conventional  algebras.

\vspace*{2mm}
\noindent \textit{Proof:}
Let $X_{k} \subset X$ be the two--dimensional subspaces
spanned by the symplectic pairs $(f_k,g_k)$, let
$\sigma_{k} \doteq \sigma \rest X_{k} \times  X_{k}$
and let $(\pi_k, \Hs_k)$ be regular irreducible representations
of $\Rs(X_k,\sigma_k)$, $k = 1, \dots N$.
Then $\pi_0 \doteq \pi_1 \otimes \cdots \otimes \pi_N$
defines an irreducible representation of the C*--tensor product
$\Rs(X_1,\sigma_1) \otimes \cdots \otimes \Rs(X_N,\sigma_N)$
on the Hilbert space
$\Hs_0 \doteq \Hs_1 \otimes \cdots \otimes \Hs_N$.
It extends by regularity to the Weyl algebra
$ \WA \simeq
\Ws(X_1,\sigma_1) \otimes \cdots \otimes \Ws(X_N,\sigma_N)$
and hence to a regular representation of $\RA$, cf.\
Proposition \ref{one-one}.

Disregarding tensor factors of $\un$ one has
$H_{0 k} \doteq
(i \mu \un - \frac{1}{2m_k} P_k^2 - \frac{m_k \omega_k^2}{2} \, Q_k^2)^{-1}
\in \pi_k(\Rs(X_k,\sigma_k))$, $k = 1, \dots N$. If~$\omega_k > 0$
this follows from the fact that the resolvent of the
harmonic oscillator Hamiltonian is a compact operator
and hence belongs to the compact ideal of
$\pi_k(\Rs(X_k,\sigma_k))$, cf.\ Proposition~\ref{compactideal}.
If $\omega_k = 0$ one resorts to the fact that the
abelian C*--algebra generated by the resolvents
\mbox{$(i\lambda \un  - P_k)^{-1}$}, $\lambda \in \RRO$
coincides with $C_0(P_k)$, the algebra of all
continuous functions of~$P_k$ vanishing at infinity.
Hence $C_0(P_k) \subset \pi_k(\Rs(X_k,\sigma_k))$
and since $(i \mu \un - \frac{1}{2m_k} P_k^2)^{-1} \in
C_0(P_k)$ the preceding statement holds also for
$\omega_k = 0$. 

Now $C_0({\RR}_+{}^{\! \! \! N}) = C_0({\RR}_+) 
\overbrace{\otimes \cdots \otimes}^N  C_0({\RR}_+)$ and 
$u_1, \dots, u_N \mapsto (i \mu - u_1 
\dots - u_N)^{-1}$ is an element of $C_0({\RR}_+{}^{\! \! \! N})$. 
Since the resolvents of the 
positive self--adjoint  operators $H_{0 k}$ generate
the abelian C*--algebras $C_0(H_{0 k})$, $k=1, \dots , N$, 
it follows from continuous functional calculus that 
$(i \mu \un - H_0)^{-1}=\big( i \mu \un - H_{0 1} \dots - H_{0 N} \big)^{-1}
\in C_0(H_{0 1}) \otimes \cdots \otimes C_0(H_{0 N}) \subset 
\pi_0(\RA)$. 

Similarly,
for the interaction potentials one uses the
fact that the abelian C*--algebras generated by the resolvents
\mbox{$(i \lambda \un - (Q_k - Q_l))^{-1}$}, $\lambda \in \RRO$ coincide
with $C_0(Q_k - Q_l)$. So as $V_{k l}\in C_0(\RR)$, one also has that
$$
V = \sum_{1 \leq k < l \leq N} \, V_{k l}(Q_k - Q_l)
\in \pi_0(\RA).
$$
In summary one gets
$(1 - (i \mu \un - H_0)^{-1}V) \in \pi_0(\RA)$
and the inverse of this operator exists if
$|\mu| > \|V\|$. Hence
$(i \mu \un - H)^{-1} = (1 - (i \mu \un - H_0)^{-1}V)^{-1}
(i \mu \un - H_0)^{-1} \in \pi_0(\RA)$
for such $\mu$. The statement for arbitrary $\mu \in \RRO$
then follows from the resolvent equation for $H$,
completing the proof.

As a matter of fact, the preceding proposition holds for a much
larger class of interaction potentials, including discontinuous 
ones. It does not hold, however, for certain physically
inappropriate Hamiltonians such as that of the
anti--harmonic oscillator \cite[Prop.~6.3]{BuGr1}. The
characterization of all Hamiltonians which are affiliated
with resolvent algebras is an interesting open problem.

We turn now to the analysis of the dynamics induced by the
Hamiltonians given above. The exponentials of the quadratic
Hamiltonians $H_0$ induce symplectic transformations, so one has
$\Ad{e^{itH_0}} (\pi_0(\RA)) = \pi_0(\RA)$ for $t \in \RR$.
For the proof that the resolvent algebra is also stable
under the adjoint action of the interacting dynamics the
crucial step consists of showing that the cocycles
$\Gamma(t) = e^{itH} e^{-itH_0}$ are elements of
$\pi_0(\RA)$. Putting \mbox{$V(t) = \Ad{e^{itH_0}}(V)$} one
can present the cocycles in the familiar form of a Dyson series
$$
\Gamma(t) = 1 + \sum_{n=1}^\infty \, i^n
\int_0^t \! dt_1 \int_0^{t_1} \! dt_2 \dots \int_0^{t_{n-1}} \! dt_n \,
V(t_n) \cdots V(t_1)
$$
and this series converges absolutely in norm since the
operators $V(t)$ are uniformly bounded. Moreover, the functions
$t \mapsto V(t)$ have values in the algebra $\pi_0(\RA)$;
but since they are only continuous in the strong operator
topology it is not clear from the outset that their
integrals, defined in this topology, are still contained
in this algebra. Here
again the specific structure of the resolvent algebra
matters. It allows to establish the desired result.

\begin{proposition} \label{dynamics}
Let $H$ be the Hamiltonian defined above. Then
$$
\Ad{e^{itH}} (\pi_0(\RA)) = \pi_0(\RA) \, , \quad t \in \RR \, .
$$
\end{proposition}

\vspace*{2mm}
\noindent \textbf{Remark:} Since $\pi_0$ is faithful
it follows from this result that
$\alpha_t \doteq \pi_0^{-1} \Ad{e^{itH}}  \pi_0$, $\, t \in \RR$
defines a one--parameter group of automorphisms of $\RA$.
It should be noted, however, that its action is not
continuous in the strong (pointwise norm) topology of $\RA$.

\vspace*{2mm}
\noindent \textit{Proof:}
Let $k,l \in 1, \dots , N$ be different numbers,
let $(f_k,g_k)$ and $(f_l, g_l)$ be symplectic pairs
as in the proof of the preceding proposition and let
$X_{k l} \subset X$ be the space spanned by
$h_{k l}(t) \doteq ((\cos{\omega_kt}) \, g_k - (\cos{\omega_lt}) \, g_l
+ (\sin{\omega_kt})/{m_k \omega_k} \, f_k -
(\sin{\omega_lt})/{m_l \omega_l} \, f_l)
$, $t \in \RR$, where we stipulate 
$(\sin{\omega t})/\omega = t$ if $\omega = 0$. 
This space is non--degenerate and, depending
on the masses and frequencies, either two or four dimensional.
We  put $\sigma_{k l} \doteq \sigma \rest X_{k l} \times X_{k l}$.
Let \mbox{$V_{k l}(t) \doteq \Ad{e^{itH_0}} (V_{k l}(Q_k - Q_l)) $}, where
$V_{k l}(Q_k-Q_l)$ is any one of the
two--body potentials contributing to $V$. Then, for any $t \in \RR$,
$$ V_{k l}(t)
=  V_{k l}((\cos{\omega_kt}) \, Q_k - (\cos{\omega_lt}) \, Q_l
+ (\sin{\omega_kt})/{m_k \omega_k} \, P_k - 
(\sin{\omega_lt})/m_l \omega_l \, P_l)
\in \pi_0(\Rs(X_{k l},\sigma_{k l})) \, .
$$
Now the function $s_1, \dots s_d \mapsto V_{k l}(s_1) \cdots V_{k l}(s_d)$
is continuous in the strong operator
topology and, for almost all $s_1, \dots s_d$,
an element of the compact ideal of $\pi_0(\Rs(X_{k l},\sigma_{k l}))$,
provided $d \geq \dim(X_{k l})$.
The latter assertion follows from the
fact that $V_{k l}(s)$ is, for given $s$, an element of the abelian C*--algebra
generated by the resolvents $\pi_0(R(\lambda,h_{k l}(s)))$,
$\lambda \in \RRO$ and that the compact ideal coincides with
the principal ideal of  $\pi_0(\Rs(X_{k l},\sigma_{k l}))$
generated by $\pi_0(R(\lambda_1,h_1) \cdots R(\lambda_d,h_d))$
for any choice of $\lambda_1, \dots \lambda_d \in \RRO$ and of
elements $h_1, \dots h_d \in X_{k l}$ which span $X_{k l}$ \cite{BuGr5}.
It is then clear that
$\big( \int_0^t \! ds \, V_{k l }(s) \big)^d =
\int_0^t \! ds_1 \cdots \int_0^t \! ds_d
\, V_{k l}(s_1) \cdots V_{k l}(s_d)$
is contained in the compact ideal of $\pi_0(\Rs(X_{k l},\sigma_{k l}))$
and this is also true for the operator $\int_0^t \! ds \, V_{k l }(s)$
since it is self--adjoint. As $k,l$ were arbitrary this implies
$\int_0^t \! dt_1  V(t_1) \in \pi_0(\RA)$.

The proof that all other terms in the Dyson
series are likewise elements of $\pi_0(\RA)$ is given by induction.
Let $I_n(t) \doteq
\int_0^t \! dt_1 \int_0^{t_1} \! dt_2 \dots \int_0^{t_{n-1}} \! dt_n \,
V(t_n) \cdots V(t_1) \in \pi_0(\RA)$, $t \in \RR$; then
$I_{n + 1}(t) = \int _0^t \! dt_1 I_n(t_1) V(t_1)$, where
the integrals are defined in the strong operator topology.
Now $t \mapsto I_n(t)$ is continuous in norm, hence
$I_{n + 1}(t)$ can be approximated according to
$$
I_{n + 1}(t) = \lim_{J \rightarrow \infty} \, \sum_{j = 1}^J
 I_n(jt/J)  \int _{(j-1)t/J}^{jt/J} \! dt_1 V(t_1)  \, ,
$$
where the limit exists in the norm topology. Since
each term in this sum is an element of $\pi_0(\RA)$
according to the induction hypothesis
it follows that $I_{n + 1}(t) \in \pi_0(\RA)$.
Because of the convergence of the Dyson series this
implies $\Gamma(t) \in \pi_0(\RA)$, $t \in \RR$,
completing the proof of the statement.

\vspace*{2mm}
Having illustrated the virtues of the resolvent algebras
for finite systems we discuss now the situation for
infinite systems. There the results are far from being complete, 
though promising. For the sake of concreteness we consider
an infinite dimensional symplectic space $\Xsig$ with a
countable symplectic basis $f_k, g_k \in X$, $k \in \ZZ$.
Similarly to the case of finite systems one can analyze
the observables and dynamics  associated with $\RA$ in any
convenient faithful representation
$(\pi_0, \Hs_0)$, such as the Fock representation.

As before, we identify the self--adjoint operators fixed
by the resolvents with the momentum and position operators of
particles,
$P_k \doteq \phi_{\pi_0}(f_k)$, $Q_k \doteq \phi_{\pi_0}(g_k)$, $k \in
\ZZ$.
In view of Haag's Theorem \cite{Em} it
does not come as a surprise that global observables,
such as Hamiltonians having a unique ground state
or the particle number operator are no longer
affiliated with the resolvent algebra of such infinite
systems. In fact, one has the
following general result \cite{BuGr5}.

\begin{lemma} Let $\Xsig$ be an infinite dimensional symplectic
space, let $(\pi_0,\Hs_0)$ be a faithful irreducible
representation of $\RA$ and let $N$ be a (possibly unbounded)
self--adjoint operator on $\Hs_0$ with an isolated
eigenvalue of finite multiplicity.
Then $(i \mu \un - N)^{-1} \notin \pi_0(\RA)$ for
$\mu \in \RRO$, \ie $N$ is not affiliated with $\RA$.
\end{lemma}

Observables corresponding to finite subsystems of
the infinite system are still affiliated with $\RA$.
Relevant examples are the partial Hamiltonians of the
form  given above,
$$
H_\Lambda \doteq
 \sum_{k \in \Lambda} \, ({\textstyle \frac{1}{2m_k}} P_k^2 +
{\textstyle \frac{m_k \omega_k^2}{2}} Q_k^2) \ + \!
\sum_{k,l \, \in \, \Lambda} \, V_{k l}(Q_k - Q_l) \, ,
$$
where $\Lambda \subset \ZZ$ is any finite set.
By exactly the same arguments as in the proof of
Proposition~\ref{observables} one can show that
any such $H_\Lambda$ is affiliated with $\RA$. Clearly,
these Hamiltonians may have isolated eigenvalues,
but these have infinite multiplicity. By the preceding
arguments one can also show that the resolvent algebra
is stable under the time evolution induced by the
partial Hamiltonians. Moreover, for suitable
potentials the evolution converges to some global
dynamics in the limit $\Lambda \nearrow \ZZ$. The precise
results are as follows.

\begin{proposition}
Let $H_\Lambda$, $\Lambda \subset \ZZ$
be the partial Hamiltonians introduced above,
where $V_{k l}$ are continuous functions tending
to $0$ at infinity, $k,l \in \ZZ$.
\begin{enumerate}
\vspace*{-2mm}
\item[(a)] Then $\Ad{e^{itH_\Lambda}} \, (\pi_0(\RA)) =  \pi_0(\RA)$,
$t \in \RR$.
\vspace*{-2mm}
\item[(b)] Let $C, D$ be positive constants such that
$\|V_{k l} \| \leq C$ and $V_{k l} = 0$ for $|k - l| \geq D$, $k,l \in
\ZZ$. Then $\lim_{\Lambda \nearrow \ZZ} \, \Ad{e^{itH_\Lambda}} $, $t \in
\RR$ exists  pointwise on $\pi_0(\RA)$ in the norm topology.
\end{enumerate}
\end{proposition}

\noindent The proof of this statement is given in \cite{BuGr5}.
It generalizes the results on a class of models describing
particles which are confined to
the points of a one--dimensional lattice by a harmonic pinning
potential and interact with their nearest neighbors \cite{BuGr1}.
In the present more general form it also has applications to other models of
physical interest. These results provide evidence
to the effect that the resolvent algebras are an expedient
framework also for the discussion of the dynamics of infinite
systems. Yet a full assessment of their power for the treatment
of such systems requires further analysis.

\section{Conclusions}
\setcounter{equation}{0}

In the present survey we have outlined some recent 
structural results and instructive applications of
the theory of resolvent algebras. These algebras
are built from the resolvents of the canonical
operators in quantum theory and their algebraic relations
encode the basic kinematical features of quantum
systems just as well as the Weyl algebras. But, as we have shown,
the novel approach cures several shortcomings of this
traditional algebraic setting.

The resolvent algebras comply with the condition that
kinematical algebras of quantum systems must have
ideals if they are to carry various dynamics of
physical interest. This requirement can easily be inferred
from the preceding arguments in case of a single
particle: there the cocycles $\Gamma(t) = e^{itH} e^{-itH_0}$
appearing in the interaction picture have the
property that the differences $(\Gamma(t) - 1)$ are
compact operators for generic interaction potentials. Hence
$(e^{itH} W e^{-itH}  - e^{itH_0} W e^{-itH_0})$ is a compact
operator for any choice of bounded operator $W$.
It is then clear that any unital \mbox{C*--algebra} which is stable under
the action of these dynamics must contain compact operators
and consequently have ideals.

The resolvent algebras, respectively their subalgebras
corresponding to finite subsystems, contain these ideals
from the outset. As we have demonstrated by several physically significant
examples, the ideals play a substantial role in the
construction of dynamics of finite and infinite
quantum systems. For they accommodate the terms in
the Dyson expansion of the cocycles resulting from the
interaction picture and
thereby entail the stability of the resolvent algebras
under the action of the perturbed dynamics.
In order to cover a wider class of models it would, however,
be desirable to invent some more direct
argument, avoiding this expansion and the ensuing
questions of convergence.

The ideals of the resolvent algebras also play a prominent
role in their classification. The nesting of primitive ideals
encodes precise information about the size of the
underlying quantum system, \ie its dimension. It
is a complete algebraic invariant in the finite dimensional case.
There is also a sharp algebraic distinction between finite
and infinite quantum systems in terms of their minimal ideals.
In either case the resolvent algebras have comfortable
algebraic properties: they are nuclear, thereby
allowing to form unambiguously tensor products with other
algebras which plays a role in the discussion of coupled
systems.

In company with the resolvents of the canonical
operators all their continuous functions
vanishing at infinity are contained in the resolvent
algebras. This feature ensures, as we have shown,
that many operators
of physical interest are affiliated with the resolvent algebras.
It also implies that these algebras contain
multiplicative mollifiers for unbounded operators which
appear in the algebraic treatment of supersymmetric models
\cite{BuGr4} or of constraint systems~\mbox{\cite{BuGr1,Co}}.
Thus the resolvent algebras provide in many respects a
natural and convenient mathematical setting for the discussion
of finite and infinite quantum systems.

\end{document}